\documentclass{webofc}
\usepackage[varg]{txfonts}   
\usepackage{bm}
\begin{document}
\title{Charmonium in electromagnetic and vortical fields}
%
%

\author{\firstname{Jiaxing} \lastname{Zhao}\inst{1}\fnsep\thanks{\email{zhao-jx15@tsinghua.org.cn}} \and
 \firstname{Shile} \lastname{Chen}\inst{1} \and
        \firstname{Pengfei} \lastname{Zhuang}\inst{1}
}
\institute{Physics Department, Tsinghua University, Beijing 100084, China}

\abstract{%
Due to larger mass and earlier production, heavy quark(quarkonium) can be sensitive probes to investigate the  fast decaying electromagnetic and vortical fields produced in heavy-ion collisions.
The non-relativistic Schr\"odinger-like equation for heavy quarks under strong electromagnetic fields in the rotating frame is deduced and used to construct the two-body equation for the charmonium system. The effective potential between charm and anti-charm becomes anisotropic in electromagnetic and vortical fields, especially along the direction of the Lorentz force. The vorticity will affect this asymmetry property largely and catalyze the transition from strong interaction dominant bound state to electromagnetic and vortical interaction controlled anisotropic bound state. It is possible to be realized in high-energy nuclear collisions. }
\maketitle
%
Many theoretical studies show the strongest electromagnetic fields generated in non-central relativistic heavy-ion collisions~\cite{Kharzeev:2007jp,Skokov:2009qp,Voronyuk:2011jd,Deng:2012pc,Tuchin:2013ie}. The maximum of the magnetic field can reach $5m_{\pi}^2$ in semi-central Au+Au collisions at top RHIC energy and almost $70m_\pi^2$ in semi-central Pb+Pb collisions at LHC energies~\cite{Deng:2012pc,Tuchin:2013ie}, where $m_\pi$ is the pion mass. In the meantime, there is a nonzero total angular momentum $J\propto b\sqrt{s_{NN}}$($b$ is the impact parameter) carried by the system of two colliding nuclei in non-central heavy-ion collisions. Although most of this total angular momentum is carried away by the spectators, there is still a sizable fraction that remains in the created quark-gluon plasma (QGP) and induces a nonzero rotational motion of QGP~\cite{Liang:2004ph,Becattini:2007sr,Jiang:2016woz}. The related global polarization of $\Lambda$ hyperons in relativistic heavy-ion collisions is measured by the STAR Collaboration~\cite{STAR:2017ckg}. It shows that the average vorticity of the QGP reaches $\omega\sim10^{21}/s$ and it's the most vortical fluid in nature. Both electromagnetic and vorticity fields decay fast with the expansion of the hot medium~\cite{Deng:2012pc,Tuchin:2013ie,Jiang:2016woz}. One needs to find a sensitive probe to characterize the electromagnetic and vortical fields. Heavy quark(quarkonium) is one of the most sensitive and effective probes due to the following three reasons: Heavy quark mass is much large than QCD cutoff, $m_c, m_b\gg \Lambda_{QCD}$, their production can be well described by perturbative QCD; Heavy quark mass is much larger than the typical temperature of the hot medium, their mass does not change in the hot medium, and the number is conserved during the evolution; Heavy quarks are produced at very early stage with the formation time $\tau_c\sim1/2m_c\approx 0.06$ fm/c for charm quark and $\tau_b\approx 0.02$ fm/c for bottom quark, they can feel the strongest electromagnetic and vortical fields. 

So far, many interesting topics related to heavy quark(quarkonium) in the electromagnetic and vortical fields emerge, see review paper~\cite{Zhao:2020jqu}. Such as, photoproduction heavy flavor mesons and quarkonium at peripheral and ultra-peripheral heavy-ion collisions~\cite{Zha:2018ytv,Shi:2017qep}, which has been observed in the experiment~\cite{ALICE:2015mzu}. The heavy quark in electromagnetic and vortical fields can generate large directed flow ($v_1$) comparing with light flavor hadrons~\cite{Chatterjee:2017ahy,Das:2016cwd,STAR:2019clv,ALICE:2019sgg}. The charmonium states will get non-collective elliptic flow ($v_2$) in the strong magnetic field~\cite{Guo:2015nsa}. In the meantime, the directed flow of charmonium states is much more sensitive to the tilted initial energy profile which is controlled by the magnitude of the global vorticity~\cite{Chen:2019qzx}. The static properties,  eg. mass and shape, of open/closed heavy flavor states will be changed in electromagnetic fields~\cite{Marasinghe:2011bt,Alford:2013jva,Cho:2014exa,Iwasaki:2021nrz}. Following these studies, we investigate charmonia in both electromagnetic and vortical fields~\cite{Chen:2020xsr}.

For heavy flavor, there exists a hierarchy of scales: $m_Q\gg m_Qv \gg m_Qv^2$. Integrating out the degrees of freedom which larger than $m_Q$ and $m_Qv$ sequentially in the QCD Lagrangian, one can get the non-relativistic versions, NRQCD and pNRQCD~\cite{Caswell:1985ui,Brambilla:1999xf}. And a potential model is derived by neglecting the interaction between color-singlet and color-octet states in the pNRQCD framework~\cite{Brambilla:1999xf}. In this case, one can employ the Schr\"odinger equation to study the properties of heavy quarkonia. The Schr\"odinger equation has been successfully applied to open and closed heavy flavors in vacuum and at finite temperature, see the review~\cite{Zhao:2020jqu}. 
Heavy quarks in the rotating medium under the electromagnetic fields can be described by the Dirac equation in the rotating frame,
\begin{equation}
\left[i \gamma^\mu (\partial_\mu+iq A_\mu+\Gamma_{\mu})-m\right]\Psi=0,
\end{equation} 
where $\Gamma_{\mu}$ is the affine connection, $A_\mu(x)$ the electromagnetic field in the rotating frame, $q$ the charge of the quark~\cite{Chen:2015hfc}. 
It's direct to calculate the electromagnetic fields in the laboratory frame in heavy-ion collisions. One needs to make a transformation for the electromagnetic field between the rotating frame and laboratory frame.
In our study, we consider a constant magnetic field along the rotating axis. Taking the symmetric gauge, the electromagnetic potential can be expressed as $A'_a=(A_0', -By'/2, Bx'/2, 0)$ with the spacetime coordinates $(t',x',y',z')$ in the laboratory frame. The electromagnetic potential in the rotating frame can be obtained via $A_\mu=(A_0-(x^2+y^2)B\omega/2, -By/2, Bx/2, 0)$ with the coordinates $(t,x,y,z)$ in the rotating frame. In the non-relativistic limit and to the first order of $1/m$, we obtain the familiar one-body Schr\"odinger equation,
\begin{equation}
\left[{({\bf p}-q{\bm A})^2\over 2m}-{q\over m}{\bf B}\cdot{\bf s}-qA_0-{\bm \omega}\cdot ({\bf s}+{\bf x}\times {\bf p})\right]\psi=E\psi.
\end{equation}
We can see clearly that there are no coupling terms between electromagnetic and vortical fields, which means that rotation does not induce any electromagnetic effect. It is necessary to emphasize that, the coupling term emerges if the electromagnetic fields are defined directly in the rotating frame~\cite{Chen:2020xsr}. 

We construct a two-body Schr\"odinger equation to describe a charmonium system in rotational medium and under the electromagnetic fields,
\begin{equation}
\left[{({\bf p}_a-q_a{\bm A}_a)^2\over 2m_a}+{({\bf p}_b-q_b{\bm A}_b)^2\over 2m_b}-q_a{\bf E}\cdot{\bf x}_a-q_b{\bf E}\cdot{\bf x}_b-{\bm \omega}\cdot ({\bf j}_a+{\bf j}_b)- {\bm \mu}\cdot{\bf B}+V\right]\psi=E\psi,
\label{eq.twoSchrodingereq}
\end{equation}
where ${\bf j}_i={\bf s}_i+{\bf x}_i\times {\bf p}_i$ is the total angular momentum which includes spin and orbital part,
$-{\bm \mu}\cdot{\bf B}$ is the spin magnetic moment-magnetic field coupling term, and $m_a=m_b=m$ and $q_a=-q_b=q$ are the heavy quark mass and charge. The interaction potential $V$ between quark and antiquark contains spin-spin interaction and spin-independent part, the later becomes the Cornell potential in vacuum. When removing the electromagnetic(vorticity) related terms, we can get the wave equation to study quarkonium in pure vortical(electromagnetic) field. Before solving Eq.~\ref{eq.twoSchrodingereq}, we first transform the single particle coordinates and momenta to the center-of-mass and relative ones, ${\bf R}=({\bf x}_a +{\bf x}_b)/2$, ${\bf r}={\bf x}_a-{\bf x}_b$, ${\bf P}={\bf p}_a+{\bf p}_b$ and ${\bf p}=({\bf p}_a-{\bf p}_b)/2$. Different from the case with only electromagnetic fields where the total wave function can always be separated into a center-of-mass and a relative part with the help of the conserved pseudo-momentum~\cite{Alford:2013jva,Chen:2020xsr}, it's no longer to find a conserved quantity to separated the two parts in a rotational field.  
However, considering the fact that the strength of the magnetic field in relativistic heavy-ion collisions can reach $eB\approx70m_\pi^2$, which is much larger than the vortical field $m_Q\omega\approx m_\pi^2$ for charm quarks, we can choose the electromagnetic related parts as the main part and take the rotational related terms as perturbations. Therefore, the Hamiltonian can be expressed as,
\begin{eqnarray}
H&=&H_{\text{EM}}+H', \nonumber\\
H_{\text{EM}}&\equiv&\frac{{\bf P}_{ps}^2}{4m}+{{\bf p}^2\over m}-{\bm \mu}\cdot{\bf B}-q{\bf E}\cdot{\bf r}-{q\over 2m}({\bf P}_{ps}\times{\bf B})\cdot{\bf r}+{q^2\over 4m}({\bf B}\times {\bf r})^2+V, \nonumber\\
H'&\equiv&-{\bm \omega}\cdot ({\bf R}\times {\bf P}_{ps})-{\bm \omega}\cdot({\bf l}+{\bf s})+{q\over 2}{\bm \omega}\cdot ({\bf R}\times ({\bf B}\times {\bf r})),
\end{eqnarray}
where ${\bf P}_{ps}={\bf P}+q{\bf B}\times {\bf r}/2$ is the pseudo-momentum which is the same as the case in pure electromagnetic fields. The contribution from the perturbation $H'$ can be calculated through the standard method in quantum mechanics. The perturbation $H'$ is relevant to the center-of-mass coordinate $\bf{R}$, it means the perturbative correction depends on the distance from quarkonium to the rotating axis. That's the nature of particle in a vortical field, and we can take the average value $\langle {\bf R}\rangle$ in the calculation. The details about numerical calculation and results are shown in~\cite{Chen:2020xsr}. Here, we just show the effective potential in the two-body Schr\"odinger equation which can give us an intuitive understanding.
\begin{figure}[h]
\centering
\includegraphics[width=3.5cm]{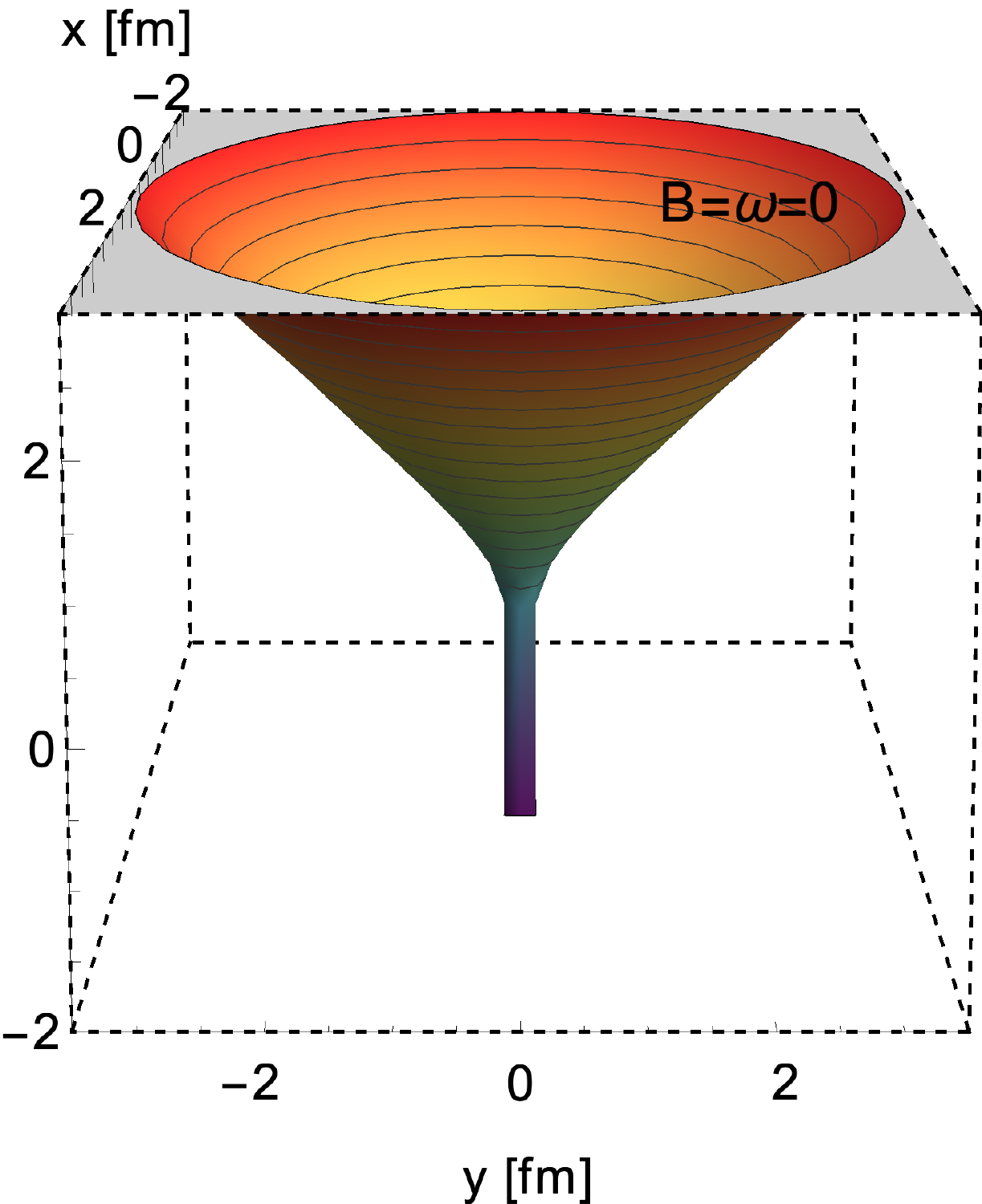} ~   \includegraphics[width=3.5cm]{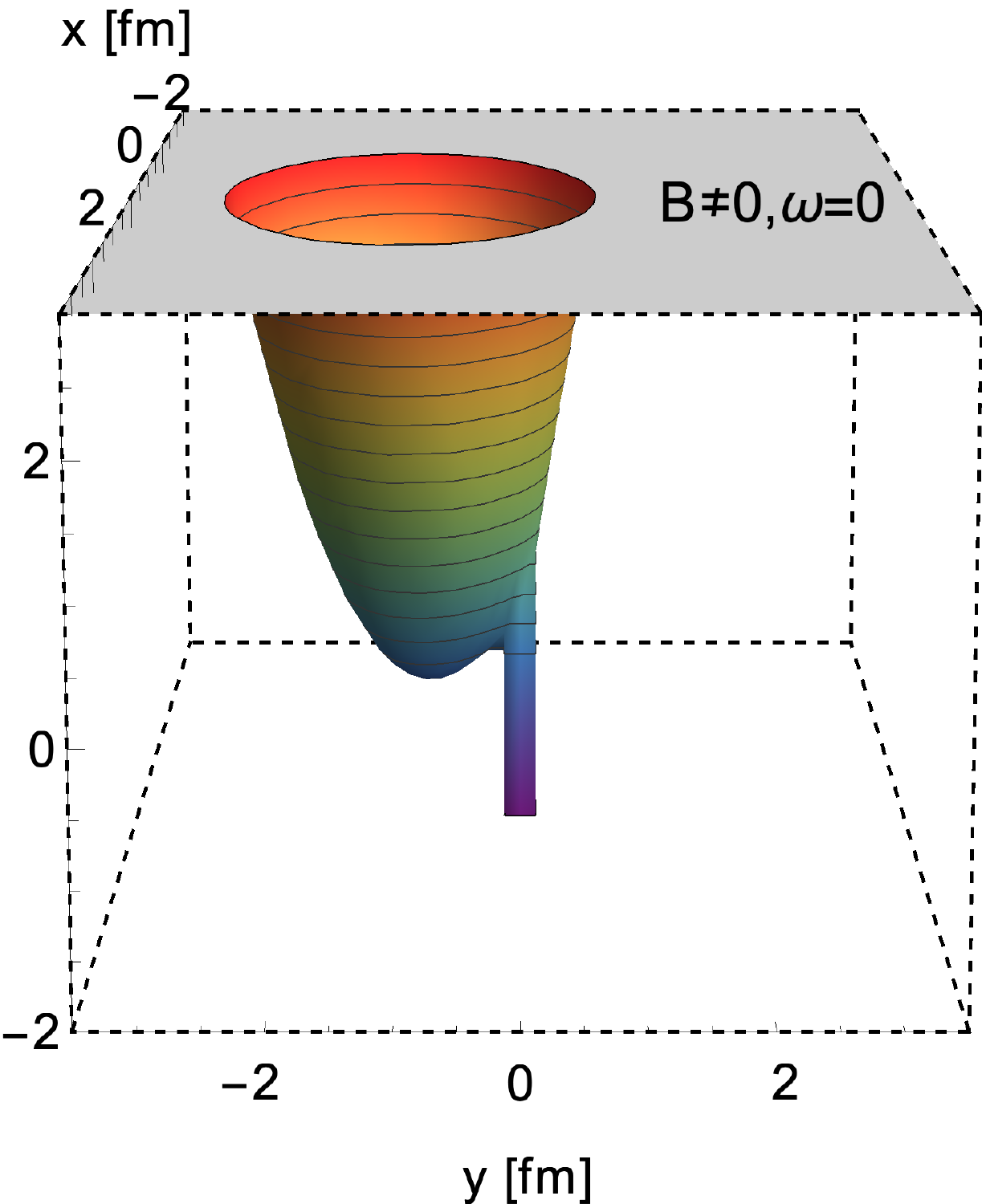} ~ \includegraphics[width=3.5cm]{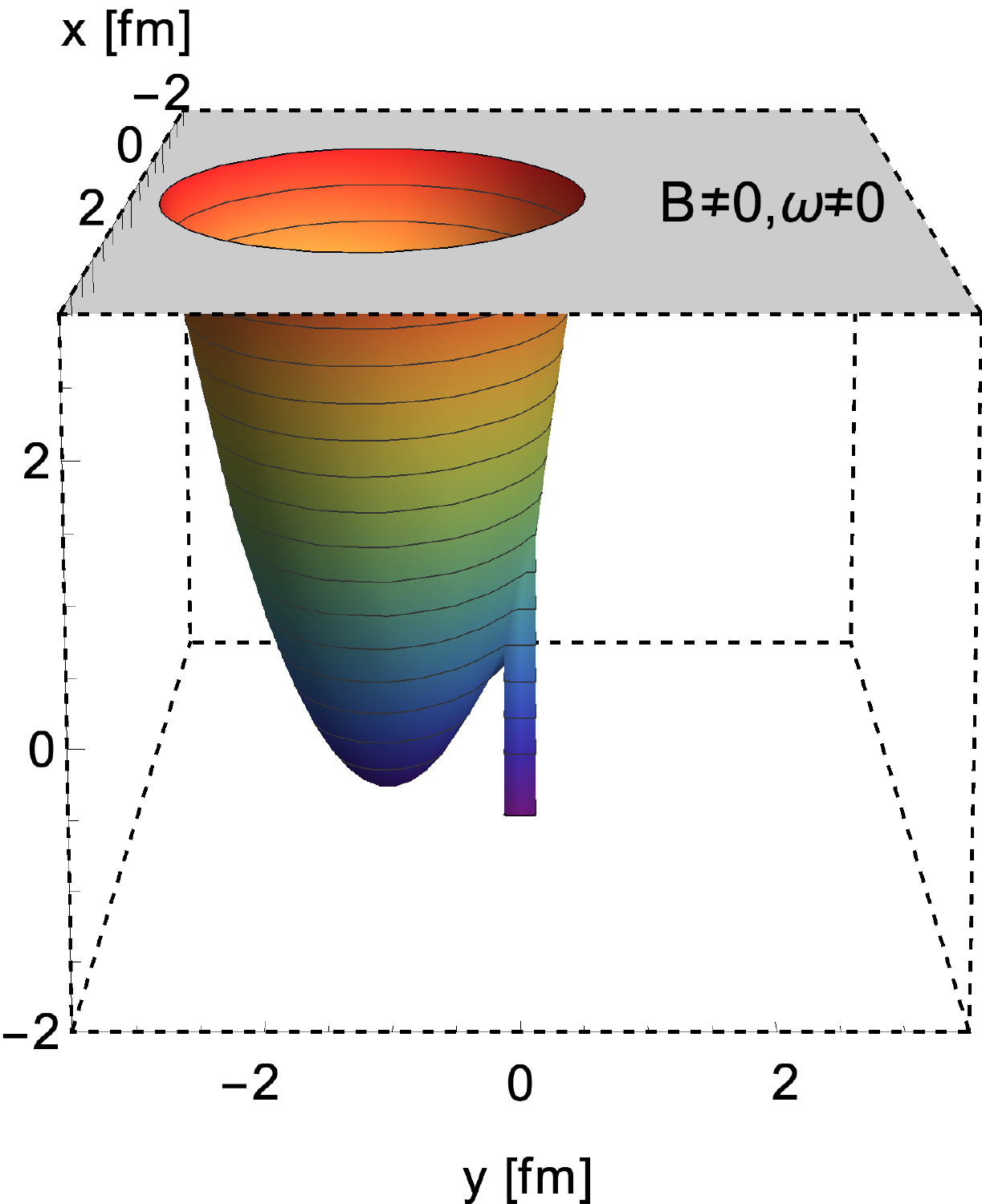}
\caption{Spin-independent vacuum interaction potential between charm and anti-charm quark in $x-y$ plane with $z=0$ with/without magnetic and/or vortical fields. The unit of potential is GeV.}
\label{fig-1} 
\end{figure}
Without magnetic field the potential(Cornell potential in vacuum) is a symmetric function in the $x-y$ plane, but the symmetry is broken when the magnetic field is turned on. When the magnetic field is strong enough, a new potential well forms and the $c\bar c$ bound state may tunnel from the strong interaction well to the magnetic well, as shown in Fig.~\ref{fig-1}. This is also known as Lorentz ionization~\cite{Marasinghe:2011bt}. The vortical field will deepen(shallow) the magnetic well. The vorticity will catalyze the transition from strongly to electromagnetically and rotationally interacting bound state. 

In summary, we derived the non-relativistic wave equation under strong electromagnetic fields in the rotating frame. We constructed the two-body Schr\"odinger equation to investigated the static properties of charmonium states in strong electromagnetic and vortical fields. Since one usually defines the electromagnetic fields in laboratory frame in heavy-ion collisions, there are no coupling terms between the magnetic field and vortical field in Dirac and Schr\"odinger equation. The effective potential between charm and anti-charm becomes anisotropic in electromagnetic and vortical fields, especially along the direction of the Lorentz force. And vorticity will increase(decrease) this asymmetry property. With increasing fields, the $c\bar c$ bound state which is confined by strong interaction in vacuum is gradually tunneled to the electromagnetic and vortical field controlled anisotropic state. It is possible to be realized in high-energy nuclear collisions. These anisotropic charmonium states will have very different dynamic dissociation in the quark-gluon medium.

{\bf Acknowledgement}: The work is supported by Guangdong Major Project of Basic and Applied Basic Research No. 2020B0301030008 and the NSFC under grant Nos. 11890712, 12075129, and 12047535.

%
%

\end{document}